\documentclass[prl, twocolumn, showpacs, superscriptaddress]{revtex4}%
\usepackage{amsfonts,amsmath,amssymb,graphicx}

\begin{document}
\title{Bohmian Mechanics and Quantum Field Theory}

\author{Detlef D\"urr}
\email{duerr@mathematik.uni-muenchen.de}
\affiliation{Mathematisches Institut der Universit\"{a}t
  M\"{u}nchen, Theresienstra{\ss}e 39, 80333 M\"{u}nchen, Germany}

\author{Sheldon Goldstein}
\email{oldstein@math.rutgers.edu}
\affiliation{Departments of Mathematics and Physics - Hill Center,
  Rutgers, The State University of New Jersey, 110 Frelinghuysen Road,
  Piscataway, NJ 08854-8019, USA}

\author{Roderich Tumulka}
\email{tumulka@mathematik.uni-muenchen.de}

\author{Nino Zangh\`\i}
\email{zanghi@ge.infn.it}
\affiliation{Dipartimento di Fisica dell'Universit\`a di Genova and
  INFN sezione di Genova, Via Dodecaneso 33, 16146 Genova, Italy}

\date{August 23, 2004}

%
\newcommand{\RRR}{\mathbb{R}}
\newcommand{\Laplace}{\Delta}
\newcommand{\I}{i}
\renewcommand{\Im}{\mathrm{Im}}
\renewcommand{\Re}{\mathrm{Re}}
\newcommand{\Q}{\mathcal{Q}}
\newcommand{\conf}{\Q}
\renewcommand{\H}{\mathcal{H}}
\newcommand{\Hilbert}{\H}
\newcommand{\pov}{{P}}
\renewcommand{\sp}[2]{\langle #1 | #2 \rangle}
\newcommand{\valpha}{{\boldsymbol \alpha}}
\newcommand{\vx}{{\boldsymbol x}}

\begin{abstract}
We discuss a recently proposed extension of Bohmian mechanics to
quantum field theory.  For more or less any regularized quantum field
theory there is a corresponding theory of particle motion, which in
particular ascribes trajectories to the electrons or whatever sort of
particles the quantum field theory is about.  Corresponding to the
nonconservation of the particle number operator in the quantum field
theory, the theory describes explicit creation and annihilation
events: the world lines for the particles can begin and end. 
\end{abstract}
\maketitle

Despite the uncertainty principle, the predictions
of nonrelativistic quantum mechanics permit particles to have precise
positions at all times.  The simplest theory demonstrating that this
is so is Bohmian mechanics \cite{Bohm52,Ischia,Stanford}; in this
theory the position of a particle cannot be known to macroscopic
observers more accurately than the $|\psi|^2$ distribution would
allow.  A frequent complaint about Bohmian mechanics is that, in the
words of Steven Weinberg \cite{weinlet}, ``it does not seem possible
to extend Bohm's version of quantum mechanics to theories in which
particles can be created and destroyed, which includes all known
relativistic quantum theories.''

To remove the grounds of the concern that such an extension may be
impossible, we show how, with (more or less) any regularized quantum
field theory (QFT), one can associate a particle theory---describing
moving particles---that is empirically equivalent to that QFT. In
particular, there is a particle theory that recovers all predictions
of regularized QED \cite{footn1}.

However, we will not attempt to achieve full Lorentz invariance; that
would lead to quite a different set of questions, orthogonal to those
with which we shall be concerned here. But we note that though the
theories we present here require a preferred reference frame, there
can be no experiment that would allow an observer to determine which
frame is the preferred one, provided the corresponding QFTs are such
that their empirical predictions are Lorentz invariant.

\begin{figure}[h]
\begin{center}
\includegraphics[width=\linewidth]{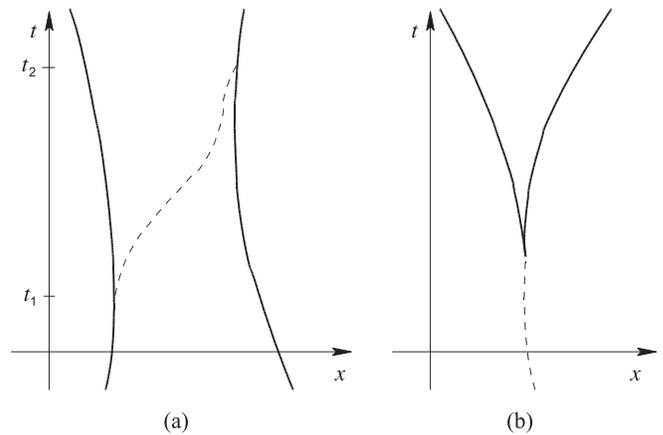}
\end{center}
\caption{Two patterns of world lines as they may arise from some
   Bell-type QFT. (a)~The world line of a photon (dashed curve) starts
   at an emission event (at time $t_1$) on the world line of an
   electron (bold curve), and ends at an absorption event (at time
   $t_2$) on the world line of another electron. (b)~An
   electron--positron pair (bold curves) is created at the end point of
   a photon world line.}
\label{figone}
\end{figure}

The theories we present are based on the work of Bell
\cite{BellBeables} and our own recent results
\cite{crea1,crea2a,crea2b}; in \cite{crea1} we study a simple model
QFT, and in \cite{crea2a,crea2b} we give a detailed account of the
mathematics needed for treating other QFTs.  While Bell replaced
physical 3-space by a lattice, we describe directly what presumably is
the continuum limit of Bell's model \cite{crea2a,crea2b,Sudbery,Vink}.
Since Bell's proposal was the first in this direction, we call these
models ``Bell-type QFTs''.  The trajectories we use as the world lines
consist of pieces of Bohmian trajectories, or similar ones.  A novel
element is that the world lines can begin and end.  This is essential
for describing processes involving particle creation or annihilation,
such as, e.g., positron--electron pair creation.  Our description of
such events is the most naive and natural one: the world line of the
particle begins at some space-time point, its creation event, and ends
at another (see figure \ref{figone}).  The models thus involve
``particle creation'' in the literal sense.

The patterns of world lines are reminiscent of Feynman diagrams, and
the possible Feynman diagrams correspond to the possible types of
world-line patterns.  Note, however, that the role of Feynman diagrams
is to aid with computing the evolution of the state vector $\Psi$,
while the world lines here are supposed to exist in addition to
$\Psi$.  Unlike Feynman diagrams, which are computational tools not to
be confused with actual particle paths, the world-line patterns of our
models are to be regarded as describing the possibilities for what
might actually happen (in a universe governed by that model).

Whatever the pattern of world lines may look like, it can be described
by a time-dependent configuration $Q_t = Q(t)$ moving in the
configuration space $\Q$ of possible positions for a variable number
of particles. In the case of a single particle species, this is the
disjoint union of the $n$-particle configuration spaces,
\begin{equation}\label{union}
   \Q = \bigcup_{n=0}^\infty \Q^{[n]} \,.
\end{equation}
Since the particles are identical, the sector $\Q^{[n]}$ is best
defined as $\RRR^{3n}$ modulo permutations, $\RRR^{3n}/S_n$.  For
simplicity, we will henceforth pretend that $\Q^{[n]}$ is simply
$\RRR^{3n}$; we discuss $\RRR^{3n}/S_n$ in \cite{crea2b}.  For several
particle species, one forms the Cartesian product of several copies of
the space \eqref{union}, one for each species.  One obtains in this
way a configuration space which is, like \eqref{union}, a union of
sectors $\Q^{[n]}$ where, however, now $n= (n_1, \ldots, n_\ell)$ is
an $\ell$-tuple of particle numbers for the $\ell$ species of
particles. For QED, for example, $\Q$ is the product of three
copies of the space \eqref{union}, corresponding to electrons,
positrons, and photons; thus, a configuration specifies the number and
positions of all electrons, positrons, and photons \cite{foot2}.

\begin{figure}[h]
\begin{center}
\includegraphics[width=\linewidth]{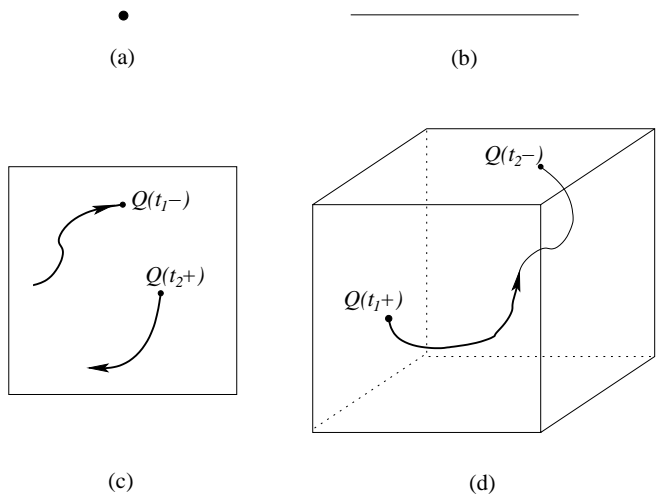}
\end{center}
\caption{Schematic representation of the configuration space of a
   variable number of particles. (a)--(d) show the sectors $\Q^{[0]}$
   through $\Q^{[3]}$.  A configurational history $Q(t)$ jumps to the
   next higher sector at each creation event, and to the next lower
   sector at each annihilation event. The history shown corresponds to
   a world-line pattern like that of figure \ref{figone}(a).}
\label{figtwo}
\end{figure}

Let us explore what $Q(t)$ looks like for a typical world line pattern
(see figure \ref{figtwo}).

\noindent $Q(t)$ will typically have discontinuities, even if there is
nothing discontinuous in the world line pattern (figure \ref{figone}),
because it jumps to a different sector at every creation or
annihilation event.  Between such events, $Q(t)$ moves smoothly within
one sector.

It is helpful to note that the bosonic Fock space can be understood as
a space of $L^2$ (i.e., square-integrable) functions on $\bigcup_n
\RRR^{3n}/S_n$.  The fermionic Fock space consists of $L^2$ functions
on $\bigcup_n \RRR^{3n}$ which are anti-symmetric under permutations.

A Bell-type QFT specifies such world-line patterns, or histories in
configuration space, by specifying three sorts of ``laws of motion'':
when to jump, where to jump, and how to move between the jumps.
Before we say more on what precisely the laws are, we elucidate one
consequence of the laws: if at $t=0$, the configuration $Q(0)$ is
chosen at random with probability distribution $|\Psi_0|^2$, then at
any later time $t$, $Q(t)$ has distribution $|\Psi_t|^2$. This
property we call \emph{equivariance}. The main consequence is that
these theories are empirically equivalent to their corresponding QFTs.
This conlusion has been explained in detail in \cite{DGZ} for Bohmian
mechanics and the predictions of nonrelativistic quantum mechanics,
and the same reasoning applies here.  It involves a law of large
numbers governing the empirical frequencies in a typical universe, and
involves the recognition that the variables that record the outcome of
an experiment are ultimately particle positions (orientations of meter
pointers, ink marks on paper, etc.).

In a Bell-type QFT, the state of a system is described by the pair
$(\Psi_t, Q_t)$, where $\Psi_t$ is an (arbitrary) vector in the
appropriate Fock space and may well involve a superposition of states
of different particle numbers. As remarked before, $\Psi_t$ can thus
be viewed as a function $\Psi_t(q)$ on the configuration space $\Q$ of
a variable number of particles. (For photons, whose position
observable is represented by a positive-operator-valued measure
(POVM), $\Psi_t$ can be represented by a wavefunction $\Psi_t(q)$
satisfying a constraint.)  $\Psi_t$ evolves according to the
appropriate Schr\"odinger equation
\begin{equation}\label{Schr}
   \I \hbar \frac{d\Psi_t}{dt} = H\Psi_t \,.
\end{equation}
Typically $H=H_0 + H_I$ is the sum of a free Hamiltonian $H_0$ and an
interaction Hamiltonian $H_I$.  It is important to appreciate that
although there is an actual particle number, defined by $N(t) = \#
Q(t) := [\text{number of entries in }Q(t)]$ or $Q(t) \in \Q^{[N(t)]}$,
$\Psi$ need not be a number eigenstate (i.e., concentrated on one
sector).  This is similar to the situation in the usual double-slit
experiment, in which the particle passes through only one slit
although the wavefunction passes through both.  And as with the
double-slit experiment, the part of the wavefunction that passes
through another sector of $\Q$ (or another slit) may well influence
the behavior of $Q(t)$ at a later time.

The laws of motion for $Q_t$ depend on $\Psi_t$ (and on
$H$). The continuous part of the motion is governed
by a first-order ordinary differential equation
\begin{equation}\label{velo}
   \frac{dQ_t}{dt} = v^{\Psi_t} (Q_t) = \Re \, \frac{\Psi_t^*(Q_t) \,
   \big( \dot{\hat{q}} \Psi_t \big)(Q_t)} { \Psi_t^*(Q_t) \,
   \Psi_t(Q_t)}
\end{equation}
\begin{equation}
   \mbox{where } \dot{\hat{q}} = \frac{d}{d\tau} \, e^{\I H_0
   \tau/\hbar} \, \hat{q} \, e^{-\I H_0 \tau/\hbar}\Big|_{\tau=0} =
   \frac{\I}{\hbar} \, \big[ H_0, \hat{q} \big]
\end{equation}
is the time derivative of the $\Q$-valued Heisenberg position operator
$\hat{q}$, evolved with $H_0$ alone. Since in the
absence of global coordinates on $\Q$, the notion of a ``$\Q$-valued
operator'' may be somewhat obscure, one should understand \eqref{velo}
as saying this: for any smooth function $f:\Q \to \RRR$,
\begin{equation}\label{vf}
   \frac{df(Q_t)}{dt} = \Re \, \frac{\Psi_t^*(Q_t) \big( \frac{\I}
   {\hbar} [H_0,\hat{f}] \Psi_t)(Q_t)} {\Psi_t^*(Q_t) \, \Psi_t(Q_t)}
\end{equation}
where $\hat{f}$ is the multiplication operator corresponding to $f$.
This expression is of the form $v^\Psi \cdot \nabla f(Q_t)$, as it
must be for defining a dynamics for $Q_t$, if the free Hamiltonian is
a differential operator of up to second order \cite{crea2b}.  The
Klein--Gordon operator is not covered by \eqref{velo} or \eqref{vf};
its treatment will be discussed in future work \cite{klein2}.  The
numerator and denominator of \eqref{velo} resp.\ \eqref{vf} involve,
when appropriate, scalar products in spin space.  One may view
$v$ as a vector field on $\Q$, and thus as consisting of one
vector field $v^{[n]}$ on every manifold $\Q^{[n]}$; it is then
$v^{[N(t)]}$ that governs the motion of $Q(t)$ in \eqref{velo}.

If $H_0$ were the Schr\"odinger operator $-\sum_{i=1}^n
\frac{\hbar^2}{2m_i} \Laplace_i +V$ of quantum
mechanics, formula \eqref{velo} would yield the velocity proposed by
Bohm in \cite{Bohm52},
\begin{equation}\label{Bohm}
   v^\Psi_i = \frac{\hbar}{m_i} \, \Im \, \frac{\Psi^* \nabla_i \Psi}
   {\Psi^* \Psi}\,, \: i=1,\ldots,n.
\end{equation}
When $H_0$ is the ``second quantization'' of a one-particle
Schr\"odinger operator, \eqref{velo} amounts to \eqref{Bohm}, with
equal masses, in every sector $\Q^{[n]}$.  Similarly, in case $H_0$ is
the second quantization of the Dirac operator $-\I c\hbar \valpha
\cdot \nabla + \beta mc^2$, \eqref{velo} says a configuration $Q(t)$
(with $N$ particles) moves according to (the $N$-particle version of)
the known variant of Bohm's velocity formula for Dirac wavefunctions
\cite{BH},
\begin{equation}\label{BohmDirac}
   v^\Psi = \frac{\Psi^* \alpha \Psi}{\Psi^* \Psi}\,c \,.
\end{equation}

The jumps are stochastic in nature, i.e., they occur at random times
and lead to random destinations. In Bell-type QFTs, God does play
dice. There are no hidden variables which would fully pre-determine
the time and destination of a jump.  (Note also that a
deterministic jump law that prescribes the time and destination of the
jump as a (smooth) function of the initial configuration would lack
sufficient randomness to be compatible with equivariance, since after
a jump from a sector with dimension $d'$ to a sector with dimension
$d>d'$ the configuration would have to belong, at any specific time,
to a $d'$-dimensional submanifold.)

The probability of jumping, within the next $dt$ seconds, to the
volume $dq$ in $\Q$, is $\sigma^\Psi (dq|Q_t) \, dt$ with
\begin{equation}\label{rates1}
   \sigma^\Psi(dq|q') = \frac{2}{\hbar} \frac{[\Im \, \Psi^*(q)  \,
   \sp{q}{H_I|q'} \, \Psi(q')]^+}{\Psi^*(q') \, \Psi(q')} \, dq \,,
\end{equation}
where $x^+ = \max(x,0)$ means the positive part of $x\in \RRR$.  Thus
the jump rate $\sigma^\Psi$ depends on the present configuration
$Q_t$, on the state vector $\Psi_t$, which has a ``guiding'' role
similar to that in Bohm's velocity law (\ref{Bohm}), and of course on
the overall setup of the QFT as encoded in the interaction Hamiltonian
$H_I$.  In \cite{crea1}, we spelled out in detail a simple example of
a Bell-type QFT.

Together, \eqref{velo} and \eqref{rates1} define a Markov process on
$\Q$.  The ``free'' part of this process, defined
by \eqref{velo}, can also be regarded as arising as follows: if $H_0$
is as usual the ``second quantization'' of a 1-particle Hamiltonian
$h$, one can construct the dynamics corresponding to $H_0$ from a
given 1-particle dynamics corresponding to $h$ (be it deterministic or
stochastic) by an algorithm that one may call the ``second
quantization'' of a Markov process \cite{crea2b}.  Moreover, this
algorithm can still be used when formula
\eqref{velo} fails to define a dynamics (in particular when $H_0$ is
the second quantized Klein--Gordon operator).

We now discuss the role of field operators (operator-valued fields on
space-time) in a theory of particles. Almost by definition, it would
seem that QFT concerns fields, and not particles.  But there is less
to this than might be expected.  The field operators do not function
as observables in QFT. It is far from clear how to actually
``observe'' them, and even if this could somehow, in some sense, be
done, it is important to bear in mind that the standard predictions of
QFT are grounded in the particle representation, not the field
representation: Experiments in high energy physics are scattering
experiments, in which what is observed is the asymptotic motion of the
outgoing particles.  Moreover, for Fermi fields---the matter
fields---the field as a whole (at a given time) could not possibly be
observable, since Fermi fields anti-commute, rather than commute, at
space-like separation. We note, though, that a theory in which
$\Psi_t$ guides an actual field can be devised, at least formally
\cite{Bohm52}.

The role of the field operators is to provide a connection, the
only connection in fact, between space-time and the abstract Hilbert
space containing the quantum states $|\Psi\rangle$, which are usually
regarded not as functions but as abstract vectors.  For our purpose,
what is crucial are the following facts that we shall explain
presently: (i)~the field operators naturally correspond to the
spatial structure provided by a projection-valued (PV) measure on
configuration space $\Q$, and (ii)~the process we have defined in this
paper can be efficiently expressed in terms of a PV measure.

Consider a PV measure $\pov$ on $\Q$ acting on $\Hilbert$: For $B
\subseteq \Q$, $\pov(B)$ means the projection to the space of states
localized in $B$. All our formulas above can be formulated in terms of
$\pov$ and $|\Psi \rangle$: \eqref{vf} becomes
\begin{equation}\label{velo2}
   \frac{df(Q_t)}{dt} = \Re \, \frac{\sp{\Psi} {\pov(dq) \frac{\I}
   {\hbar} [H_0,\hat{f}] |\Psi}} {\sp{\Psi}{\pov(dq)|\Psi}}
   \Big|_{q=Q_t}
\end{equation}
\begin{equation}
    \mbox{with } \hat{f} = \int\limits_{q\in\Q} f(q) \, \pov(dq) \,,
\end{equation}
for any smooth function $f:\Q \to \RRR$, and \eqref{rates1} becomes
\begin{equation}\label{rates2}
   \sigma^\Psi(dq|q') = \frac{2}{\hbar} \frac{[\Im \, \sp{\Psi} {\pov
   (dq) H_I \pov(dq') | \Psi} ]^+} {\sp {\Psi} {\pov(dq') |\Psi}}.
\end{equation}
Note that $\sp{\Psi} {\pov(dq) |\Psi}$ is the probability distribution
analogous to $|\Psi(q)|^2 dq$.

We now turn to (i): how we obtain the PV measure $\pov$ from the field
operators.  For the configuration space $\Q= \bigcup_n \RRR^{3n}/S_n$
of a variable number of identical particles, a configuration can be
specified by giving the number of particles $n(R)$ in every region
$R \subseteq \RRR^3$. A PV measure $\pov$ on $\Q$ is mathematically
equivalent to a family of number operators: an additive
operator-valued set function $N(R)$, $R \subseteq \RRR^3$, such that
the $N(R)$ commute pairwise and have spectra in the nonnegative
integers. Indeed, $\pov$ is the joint spectral decomposition of the
$N(R)$ \cite{crea2b}.  And the easiest way to obtain such a family of
number operators is by setting
\[
   N(R) = \int_R \phi^*(\vx) \, \phi(\vx) \, d^3\vx\, ,
\]
exploiting the canonical commutation or anti-commutation relations
for the field operators $\phi(\vx)$.  These observations suggest
that field operators are just what the doctor ordered for the
efficient construction of a theory describing the creation,
motion, and annihilation of particles.

(It is only the positive-energy one-particle states that are used for
constructing the Fock space $\Hilbert$, so that $\Hilbert$ is really a
subspace of a larger Hilbert space $\Hilbert_0$ which contains also
unphysical states (with contributions from one-particle states of
negative energy).  Since position operators may fail to map positive
energy states into positive energy states, the PV measure $\pov$ is
typically defined on $\Hilbert_0$ but not on $\Hilbert$, in which case
\eqref{velo2} and \eqref{rates2} have to be read as applying in
$\Hilbert_0$.  While $H_0$ is defined on $\Hilbert_0$, $H_I$ is
usually not and needs to be ``filled up with zeroes'', i.e.\ replaced
by $P'H_IP'$ where $P'$ is the projection $\Hilbert_0 \to \Hilbert$.)

To sum up, we have shown how the realist view which Bohmian mechanics
provides for the realm of nonrelativistic quantum mechanics can be
extended to QFT, including creation and annihilation of particles.
Those who find the all too widespread positivistic attitude in quantum
theory unsatisfactory may find these ideas helpful.  But even those
who think that Copenhagen quantum theory is just fine may find it
interesting to see how the particle picture, ubiquitous in the
pictorial lingo and heuristic intuition of QFT, can be made
consistent, internally and with the observable facts of QFT, by
introducing suitable laws of motion.

\end{document}